\documentclass[%
 reprint,
superscriptaddress,
 amsmath,amssymb,
 aps,
prb,
floatfix,
]{revtex4-2}

\usepackage{graphicx}
\usepackage{dcolumn}
\usepackage{bm}
\usepackage{hyperref}
\usepackage{xcolor}

\begin{document}

\title{$U(1)$ Gauge-Equivariant Representation Learning of Bloch States}

\author{Chengyan Zhang}
\author{Xian Xu}
\affiliation{Department of Applied Physics, Yale University, New Haven, CT, 06511, USA}
\author{Bowen Hou}
\author{Diana Y. Qiu}
\email{diana.qiu@yale.edu}
\affiliation{Department of Materials Sciences, Yale University, New Haven, CT 06511, USA}

\date{\today}

\begin{abstract}
Representation learning, or featurization, of mean-field Bloch states provides an important route for incorporating electronic information into machine-learning models of first-principles condensed-phase systems.
A challenge in representing Bloch states is the gauge redundancy of each state.
In this work, we propose to address this issue by explicitly incorporating gauge equivariance in the machine-learning framework.
We develop a $U(1)$-equivariant autoencoder that compresses the plane-wave-basis Bloch states obtained from density functional theory into a low-dimensional latent representation.
We achieve an average overlap of 0.970 between original wavefunctions and the reconstructions from a 64-dimensional latent space using a dataset of two-dimensional insulators.
We further show that $U(1)$ equivariance endows the latent space with physically meaningful structures, including smoothness and topological information.
Finally, we present a proof-of-principle application in which the latent representations are used to predict the off-diagonal elements of the $GW$ self energy matrix, and demonstrate the importance of enforcing gauge equivariance in the prediction.

\end{abstract}

\maketitle
\begin{figure*}[t]
    \centering
    \includegraphics[width=0.99\textwidth]{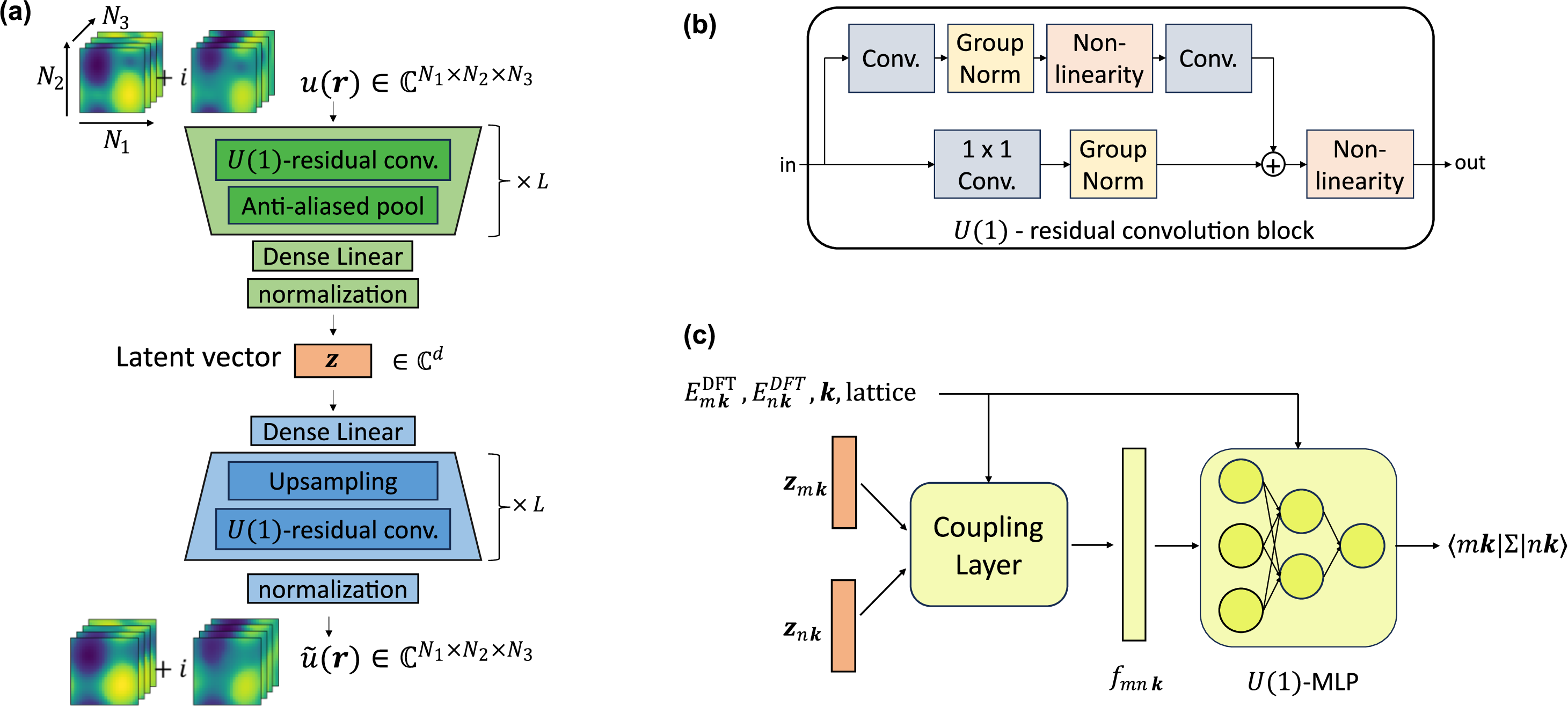}
    \caption{Neural network architecture of $U(1)$-autoencoder and a proof-of-principle prediction network.
    \textbf{(a)} Overall structure of the autoencoder. The encoder is depicted in green, and the decoder is in blue. Both the encoder and decoder contain multiple compression blocks or upsampling blocks reducing or increasing spatial dimensions.
    \textbf{(b)} Residual convolution block used in the architecture.
    \textbf{(c)} A simple multilayer perceptron architecture used to predict matrix elements from individual latent vectors. Two latent vectors are first assembled into combined features before going through a multilayer perceptron, where all operations are made $U(1)$-equivariant.
    }
    \label{fig:architecture}
\end{figure*}

\section{Introduction}
Machine learning has been widely adopted in the modeling of first-principles systems.
Models using atomic features, such as chemical species and crystal structures, have achieved substantial success in applications including interatomic potentials \cite{Behler2007generalized, Kocer2022neural, Batzner2022e3equivariant}, prediction of physical properties \cite{Schutt2018schnet, Xie2018crystala}, and acceleration of electronic structure calculations \cite{Li2022deeplearninga, Gong2023general, Zhong2023transferable}. 
By contrast, the use of electronic features in machine learning remains comparatively less explored.

Recently, there has been growing interest in using features extracted from single-particle mean-field (MF) electronic structure as inputs to machine-learning models, particularly for excited-state or post-MF targets \cite{Welborn2018transferability, Qiao2022informing, Venturella2024machinea, Venturella2025unified, Thiede2026coupled, Knosgaard2022representing, Zadoks2024spectral, Hou2024unsuperviseda, Hou2025mbformer}.
MF methods, such as Hartree-Fock theory \cite{hartree1928wave,Fock1930Naeherungsmethode,Slater1930Note} and Kohn-Sham Density Functional Theory (DFT) \cite{Hohenberg1964Inhomogeneous, Kohn1965SelfConsistent} are computationally efficient while encoding rich information about the electronic structure.
For example, the MF eigenvalues provide quantities such as the MF band structure, Fermi level, and density of states, while the MF eigen wavefunctions encode information including orbital and bonding character \cite{Marzari2012maximally}, Berry phases \cite{Xiao2010berry, Vanderbilt2018berry}, and topological properties \cite{Bansil2016colloquium}.
In addition, the MF states provide a standard starting point for \textit{ab initio} many-body perturbation theory calculations of many excited-state properties and correlated electronic structures, including the GW approximation~\cite{Hybertsen1986electron, Godby1989metalinsulator}, the Bethe-Salpeter equation (BSE) \cite{Albrecht1998initio, Benedict1998opticala, Rohlfing1998electronholea, Rohlfing2000electronhole, Onida2002electronica}, and  coupled-cluster theory \cite{Bartlett2007coupledcluster}.
These considerations make MF electronic states natural input features for machine learning models targeting post-MF and other complicated electronic properties.

A central challenge in using MF electronic information as machine learning input is the featurization of the MF wavefunctions.
The raw wavefunction $\psi(\bm{r})$ belongs to a high-dimensional functional Hilbert space, and a compact representation is required.
Several featurization strategies have recently been proposed.
In molecular systems, representations based on atomic-orbital basis are commonly used, including MF electronic matrix \cite{Welborn2018transferability, Qiao2022informing}, MF Green's function features \cite{Venturella2024machinea, Venturella2025unified}, and molecular orbital coefficients \cite{Thiede2026coupled}.
In periodic systems, the MF wavefunctions form Bloch states $\psi_{n\bm{k}}(\bm{r})=e^{i\bm{k}\cdot\bm{r}}u_{n\bm{k}}(\bm{r})$ with band index $n$, momentum $\bm{k}$, and a cell-periodic function $u_{n\bm{k}}(\bm{r})$.
Some different representations have also been explored in this setting, including operator and density of states features \cite{Knosgaard2022representing, Zadoks2024spectral}, as well as the direct representation learning of individual Bloch states \cite{Hou2024unsuperviseda, Hou2025mbformer}.

Representation learning provides a direct and flexible route to the featurization of MF wavefunctions, independent of human biases, and has recently been used to learn the long-range many-body correlations in materials at the GW and BSE levels~\cite{Hou2024unsuperviseda, Hou2025mbformer}.
In this approach, the periodic part $u_{n\bm{k}}(\bm{r})$ of a MF Bloch state under the plane-wave basis, or its representation in real space, is treated as a high-dimensional vector on a grid.
An encoding neural network consisting of an encoder ($\mathrm{enc}$) and a decoder ($\mathrm{dec}$) is then trained to compress each Bloch state into a low-dimensional latent vector and reconstruct the original wavefunction such that
\begin{align}
    &\bm{z}_{n\bm{k}}=\mathrm{enc}(u_{n\bm{k}}\big(\bm{r})\big),\label{eq:encoder} \\ 
    &\tilde{u}_{n\bm{k}}(\bm{r})=\mathrm{dec}(\bm{z}_{n\bm{k}}), \label{eq:decoder}
\end{align}
where $\bm{z}_{n\bm{k}}$ and $\tilde{u}_{n\bm{k}}$ denote the latent vector and reconstructed wavefunction, respectively.
The encoder and decoder networks are optimized in an unsupervised manner so that $\tilde{u}_{n\bm{k}}(\bm{r})$ approximates $u_{n\bm{k}}(\bm{r})$, thus forcing $\bm{z}_{n\bm{k}}$ to encode essential information contained in the original wavefunction. The latent vectors $\bm{z}_{n\bm{k}}$ are then used as input features to downstream prediction models:
\begin{equation}\label{eq:prediction}
    O=F\big(\{\bm{z}_{n\bm{k}}, E_{n\bm{k}}\}\big)
\end{equation}
for some target $O$, where, in principle, all the latent vectors $\bm{z}_{n\bm{k}}$ and MF energies $E_{n\bm{k}}$ across the whole Brillouin zone within an energy window of interest should be included in the input.
This approach has several advantages.
It is fully data-driven and automated, retains most of the information of each Bloch state, and produces a fixed-dimensional structured vector representation readily used in sophisticated downstream prediction architectures, including transformers \cite{Hou2025mbformer}.

However, a key obstacle to representation learning of individual Bloch states is the gauge redundancy.
In the absence of degeneracy, each Bloch state contains an arbitrary $U(1)$ gauge freedom:
\begin{equation}\label{eq:gaugeTransform}
    u_{n\bm{k}}(\bm{r})\mapsto e^{i\theta_{n\bm{k}}}u_{n\bm{k}}(\bm{r}),
\end{equation}
and the gauge resulting from numerical eigensolvers is generally uncontrolled.
Consequently, previous work focused on learning representations from the MF wavefunction amplitudes $|\psi_{n\bm{k}}(\bm{r})|$ only, discarding all phase information and thereby limiting predictions to gauge-invariant quantities, such as energies \cite{Hou2024unsuperviseda, Hou2025mbformer}. An alternative strategy, recently used in unsupervised learning of fractional Chern insulators, is to explicitly fix the Bloch-state gauge prior to representation learning \cite{Wu2026modelinga}.
Here, we propose an alternative approach to solve the gauge redundancy challenge in representation learning [Eq. \eqref{eq:encoder}-\eqref{eq:decoder}], while retaining the full complex wavefunction, by incorporating $U(1)$-equivariance into the neural network architecture.
Specifically, we develop a $U(1)$-equivariant autoencoder (AE) for the representation learning of Bloch states, and demonstrate that enforcing $U(1)$ equivariance yields latent representations with physically meaningful structures, including Berry curvature and band topology. Interestingly, we find that a non-equivariant autoencoder can also learn an approximate gauge structure, providing insight into how gauge equivariance can emerge implicitly from training data.
Finally, we show that retaining phase information enables the prediction of complex gauge-covariant quantities, which we demonstrate through the downstream prediction of off-diagonal matrix elements of the $GW$ self energy as a proof-of-principle.

The remainder of this paper is organized as follows.
In Sec. \ref{sec:u1ae} we formulate the $U(1)$ equivariance condition and construct the corresponding equivariant autoencoder.
In Sec. \ref{sec:properties} we discuss the physically meaningful properties of the latent representation space.
In Sec. \ref{sec:prediction} we present a proof-of-principle demonstration of the role of phase information and gauge equivariance in predicting gauge-sensitive quantities.
Finally, in Sec. \ref{sec:discussion} we discuss the relation of the present study to the broader context of gauge equivariance, and point out limitations and future directions.
\section{$U(1)$-equivariant autoencoder (AE)}\label{sec:u1ae}

We consider a machine-learning task approximating a map from a set of mean-field Bloch states to a certain output quantity.
Under a gauge transformation, the target map transforms consistently.
This gauge structure can thus be explicitly incorporated into the machine learning model as an equivariant constraint.
As indicated by Eq. \eqref{eq:encoder}-\eqref{eq:prediction}, the machine learning model consists of two stages: a representation learning stage and a downstream prediction stage.
Gauge equivariance takes two different forms in the two stages.
Let $\bm{z}_{n\bm{k}}=\text{encoder}\big(u_{n\bm{k}}(\bm{r})\big)\in\mathbb{C}^d$, where $d$ is the latent representation dimension.
Throughout this work, we neglect degeneracies and consider only the $U(1)$ gauge transformation in Eq. \eqref{eq:gaugeTransform}.
Because the encoder and decoder act on each Bloch state independently, the equivariant condition for representation learning (Eq. \eqref{eq:encoder}-\eqref{eq:decoder}) is a global equivariance:
\begin{align}
    &\mathrm{enc}\big(e^{i\theta} u(\bm{r})\big)=\rho_z(e^{i\theta})\cdot\mathrm{enc}\big(u(\bm{r})\big),\label{eq:encoderGauge}\\
    &\mathrm{dec}\big(\rho_z(e^{i\theta})\,\bm{z}\big)=e^{i\theta}\,\mathrm{dec}(\bm{z}),\label{eq:decoderGauge}
\end{align}
where $\rho_z$ is a representation of $U(1)$ in the latent space $\mathbb{C}^d$.
By contrast, the downstream prediction model (Eq. \eqref{eq:prediction}) should in principle take all the relevant Bloch states as input, each of which has a $U(1)$ redundancy. Its equivariance is therefore a local gauge equivariance:
\begin{equation}\label{eq:predGauge}
    F\big(\{\rho_z(e^{i\theta_{n\bm k}})\,\bm z_{n\bm k}, E_{n\bm k}\}\big)=\rho_O(\{e^{i\theta_{n\bm{k}}}\})\cdot F\big(\{\bm z_{n\bm k}, E_{n\bm k}\}\big),
\end{equation}
where $\rho_O$ is a representation of the gauge group consisting of all gauge transformations.

The representation $\rho_O$ is determined by the target output $O$, while $\rho_z$ should be chosen so that the condition Eq. \eqref{eq:predGauge} can be realized relatively easily.
For a physical observable $O$, such as the GW quasiparticle energies and the exciton energies, $\rho_O=1$ is the trivial representation, as observables are gauge-independent.
In this case, $\rho_z$ can be chosen to contain a trivial representation of $U(1)$, and the corresponding gauge-invariant subspace can be used as the input of the prediction model $F$, so that the equivariance condition \eqref{eq:predGauge} is satisfied trivially.
On the other hand, $\rho_O$ is nontrivial if the target $O$ is an operator matrix expressed in the Bloch-state basis, such as the self energy matrix $\langle m\bm{k}|\Sigma|n\bm{k}\rangle$:
under the gauge transformation in Eq. \eqref{eq:gaugeTransform}, each element acquires a phase $\exp[i(\theta_{n\bm{k}}-\theta_{m\bm k})]$. 
For these targets, it is natural to choose the latent space action of the gauge transformation $\rho_z$ as the defining representation $\rho_z(e^{i\theta})=e^{i\theta}$.

In this work, we focus primarily on the case where $\rho_O$ is nontrivial and $\rho_z$ is the defining representation.
This choice is motivated by two considerations.
First, previous work has achieved high accuracy in the prediction of real gauge-invariant physical observables using only amplitude information of the Bloch states \cite{Hou2024unsuperviseda, Hou2025mbformer}, and we expect phase information to be more important in complex targets that transform nontrivially under gauge transformations.
Second, linear transformations are automatically equivariant under the defining representation as $W\bm{z}e^{i\theta}=e^{i\theta}W\bm{z}$ for an arbitrary matrix $W$.
Consequently, the equivariance conditions in representation learning Eq. \eqref{eq:encoderGauge}-\eqref{eq:decoderGauge} can be realized without requiring a highly sophisticated network architecture.

\begin{figure}[!t]
    \centering
    \includegraphics[width=0.99\linewidth]{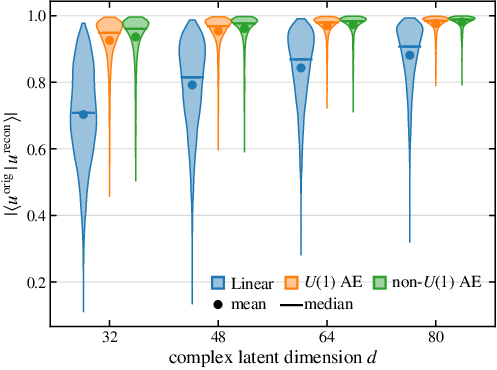}
    \caption{Overlap between original and reconstructed wavefunctions $|\langle u^{\mathrm{orig}}|u^{\mathrm{recon}}\rangle|$ on the test dataset at different latent dimensions.
    The thickness of the shaded regions shows the distribution of overlaps. The dot and the dash show the mean and median, respectively.
    For each dimension, the three columns from left to right are the linear AE, $U(1)$ AE and non-$U(1)$ AE. The latent space of the non-$U(1)$ AE is real, in which two real dimensions are counted as one complex dimension.}
    \label{fig:violin}
\end{figure}

Motivated by these considerations, we develop a $U(1)$-equivariant AE in which the latent space transforms according to the defining representation of $U(1)$, $\rho_z(e^{i\theta})=e^{i\theta}$.
The architecture consists of an encoder and a decoder, and is illustrated in Fig.~\ref{fig:architecture}(a)(b).
The periodic parts of the Bloch functions, $u(\bm{r})$, are first represented on a fixed-size real-space grid containing $N_1$, $N_2$ and $N_3$ points along the three crystal axes.
The encoder contains $L$ compression layers, each consisting of a residual convolution block \cite{He2016deep, Wu2018Group} with circular padding followed by an antialiased pooling block.
These layers compress the $N_1\times N_2\times N_3$ spatial dimensions of the wavefunctions, while creating a channel feature dimension.
A dense linear layer then maps the feature maps into a fixed-dimensional latent vector $\bm{z}\in\mathbb{C}^d$.
The decoder mirrors the encoder and reconstructs the $N_1\times N_2\times N_3$ spatial dimensions.
In this architecture, all linear, pooling, upsampling and normalization layers are inherently $U(1)$-equivariant under the defining representation.
To ensure that all nonlinear activation functions $a$ are also $U(1)$ equivariant, we choose the form
\begin{equation}\label{eq:U1activation}
    a(x)=\sigma(|x|)\frac{x}{|x|+\epsilon},\quad x\in\mathbb{C}
\end{equation}
for all activations, where $\sigma$ is an arbitrary non-linear scalar function and $\epsilon$ is a small positive number for regularization.
Further details of the neural network architecture are provided in appendix \ref{sec:app-NN-AE}.
The model will be trained in an unsupervised manner by minimizing the mean-squared reconstruction error over $N_{\mathrm{wfn}}$ wavefunctions:
\begin{equation}
    \mathcal{L}=\frac{1}{N_{\mathrm{wfn}}N_1N_2N_3}\sum_m^{N_{\text{wfn}}}\sum_{i}^{N_1N_2N_3}\big|u_m(\bm{r}_i)-\tilde{u}_m(\bm{r_i})\big|^2,
\end{equation}
where $\tilde{u}(\bm{r}_i)=\mathrm{dec}[\mathrm{enc}\big(u(\bm{r}_i)\big)]$.
We use the squared reconstruction error rather than an overlap-based loss so that the reconstruction and latent representation retain the gauge of the original wavefunction.

We emphasize that the use of a fixed-size real-space grid is a convenient implementation choice rather than an inherent limitation to our framework.
This is based on two considerations.
First, the size of the grid can be made sufficiently dense so that most of the charge is included in the representation.
Second, if the input grid size is allowed to vary, the dense linear layers can be replaced by appropriate pooling and upsampling blocks to yield a similar output.

Another two AE architectures are also considered for comparison.
The first is a linear AE, which is a special case of the $U(1)$-equivariant AE in Fig. \ref{fig:architecture}(a).
Its encoder and decoder each consist of only one dense linear layer: $\mathrm{enc}(u)=Wu$ and $\mathrm{dec}(\bm{z})=W^{\dagger}\bm{z}$.
At the global minimum trained under mean squared loss, the linear AE learns the same subspace as a principal component analysis (PCA) \cite{Baldi1989neural}.
The second architecture is a nonlinear, non-equivariant AE that ignores the $U(1)$ symmetry and the complex structure of wavefunctions.
In this architecture, each complex dimension is treated as two real dimensions, and the neural network only consists of real-valued layers.
See appendix \ref{sec:app-NN-AE} for additional details.
These two additional architectures are used to isolate the effects of nonlinearity and gauge equivariance.

Following Refs. \cite{Hou2024unsuperviseda, Hou2025mbformer}, we test the AE using a dataset of 2-dimensional insulators, and the architecture can be straightforwardly generalized to 3 dimensions.
Because 2-dimensional materials do not have translational symmetry along the out-of-plane crystal axis, we treat the $N_3$ grid points along this direction as a channel dimension, and use 2-dimensional convolution blocks to compress the $N_1\times N_2$ in-plane spatial dimensions.
The models are trained on the DFT wavefunctions of 800 materials whose structures are sampled from the C2DB dataset \cite{Haastrup2018computational, Gjerding2021recent}, with a train-validation split of 0.85:0.15.
We generate the wavefunctions using Quantum ESPRESSO \cite{Giannozzi2009quantum, Giannozzi2017advanced}.
For each material, 10 bands and a $5\times5\times 1$ k-point grid are taken, with wavefunctions represented on a $24\times 24\times 40$ real-space grid. An additional set of 34 materials is reserved as the test set.
Further details of dataset generation are provided in appendix \ref{sec:app-data}.

Fig.~\ref{fig:violin} shows the overlap between the original and reconstructed wavefunctions in the test dataset.
The $U(1)$-equivariant AE achieves an average overlap of $0.970$ with 64 complex latent dimensions and $0.976$ with 80 dimensions.
We take the latent dimension to be 64 through the rest of the paper.
We compare the three architectures of AE: the linear AE, the nonlinear $U(1)$-equivariant AE and the nonlinear non-equivariant AE.
The linear AE exhibits substantially larger reconstruction errors than the nonlinear AEs, indicating the importance of nonlinearity in compact information encoding and accurate reconstruction.
Linear compression schemes are fundamentally limited by the dimension of the subspace spanned by the vectors in the dataset.
They are thus particularly useful when the vectors of interest are mostly linearly dependent, or equivalently have a low-rank structure \cite{Hou2025datadriven}.
On the other hand, nonlinear schemes do not have this limitation and have the potential to achieve a higher compression rate and reconstruction quality in a more diverse dataset, such as the wavefunction dataset considered in this paper.
By contrast, the $U(1)$-equivariant and non-equivariant AE have similar reconstruction quality, indicating that enforcing $U(1)$ equivariance does not improve reconstruction by itself, despite other advantages shown below, and also suggesting that the AE is capable of learning $U(1)$ symmetry in reconstruction even when equivariance is not enforced.

\begin{figure}[!t]
    \centering
    \includegraphics[width=0.8\linewidth]{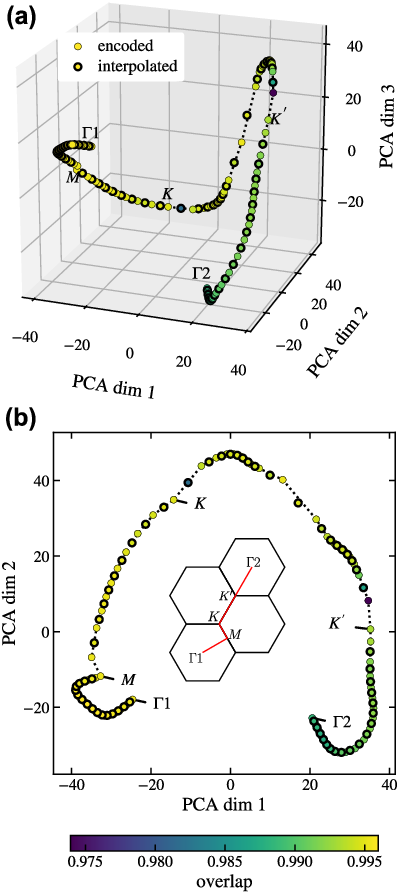}
    \caption{The 3-dimensional (a) and 2-dimensional (b) principal component analysis (PCA) visualization of the 64-dimensional latent density matrices corresponding to a path of wavefunctions in the highest valence band of monolayer $\text{MoS}_2$. The points with thin edges are directly encoded from wavefunctions, and the bold points are interpolated from neighboring points. The color of each point shows the overlap of the wavefunction reconstructed from the point and the actual wavefunction on the path.}
    \label{fig:path}
\end{figure}

\section{Properties of the Latent Representation}\label{sec:properties}
In this section, we examine the physically meaningful structures of the latent space induced by the $U(1)$ equivariance.
We emphasize that these properties follow from the general $U(1)$ equivariance and the smoothness of neural networks, and are not specific to the particular AE architecture introduced above.

\subsection{Equivariance and Smoothness}\label{sec:equivariance}

\begin{figure}[!t]
    \centering
    \includegraphics[width=0.8\linewidth]{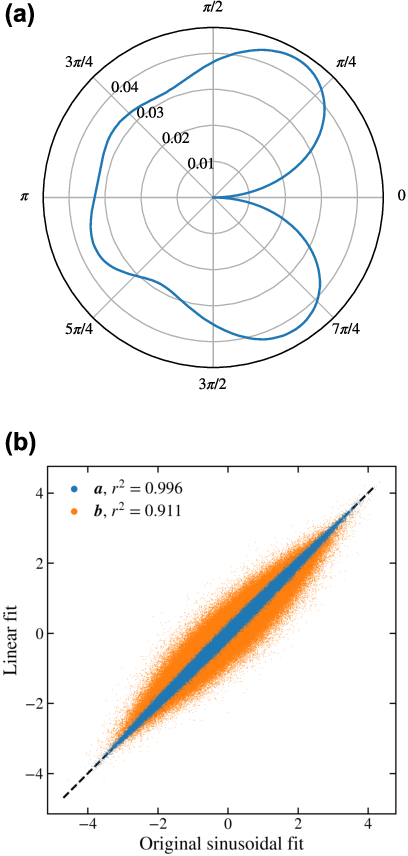}
    \caption{$U(1)$ gauge transformation $\rho_{\theta}(\bm{z})$ in the real 128-dimensional non-equivariant latent space. The transformation fits well to a sinusoidal form $\tilde{\rho}_\theta(\bm{z})=\bm{a}(\bm{z})[\cos\theta-1]+\bm{b}(\bm{z})\sin\theta+\bm{z}$.
    \textbf{(a)} Relative error between actual gauge-transformed vector $\rho_{\theta}(\bm{z})$ and the sinusoidal fit $\tilde{\rho}_\theta(\bm{z})$ as a function of $\theta$, i.e., $|\rho_\theta(\bm{z})-\tilde{\rho}_\theta(\bm{z})|\,/\,|\bm{z}|$, at a typical $\bm{z}$.
    The radial axis shows the relative error, while the angular coordinate shows $\theta$.
    \textbf{(b)} Parity plot of a linear fit to the $\bm{z}$-dependence of the sinusoidal coefficients $\bm{a}(\bm{z})$ and $\bm{b}(\bm{z})$. The horizontal axis shows the actual coefficients $\bm{a}$, $\bm{b}$ at each $\bm{z}$ extracted from sinusoidal fit, while the vertical axis shows the linear regression values of $\bm{a}$, $\bm{b}$ with respect to $\bm{z}$.}
    \label{fig:realEquivariance}
\end{figure}

The complex latent vectors $\bm{z}_{n\bm{k}}\in\mathbb{C}^d$ produced by the $U(1)$-equivariant AE are gauge equivariant by construction.
Under a $U(1)$ gauge transform $u_{n\bm{k}}(\bm{r})\mapsto e^{i\theta_{n\bm{k}}}u_{n\bm{k}}(\bm{r})$, the corresponding latent vector transforms by the same phase $\bm{z}_{n\bm{k}}\mapsto e^{i\theta_{n\bm{k}}}\bm{z}_{n\bm{k}}$.
The gauge can therefore be manipulated directly in the latent space in the same manner as that of the original wavefunction, facilitating gauge-consistent downstream applications.
For example, the latent density matrix $\bm{z}_{n\bm{k}}\bm{z}_{n\bm{k}}^\dagger$ is gauge invariant and eliminates the uncontrolled gauge adopted by the eigensolver, thereby revealing the smooth variation of the encoded states across the reciprocal space.
As an illustration, Fig.~\ref{fig:path} shows the PCA visualization of the latent density matrices obtained from a 64-dimensional complex latent space for the highest valence band of monolayer $\mathrm{MoS}_2$ along a high-symmetry path in the reciprocal space.
The latent density matrices vary smoothly along the path, consistent with the smooth variation of the underlying physical states.

To further test the smoothness of latent representation, we remove half of the points on the path in Fig. \ref{fig:path}, and use the remaining points to interpolate the removed points using linear interpolation under parallel-transport gauge.
The interpolated points are shown with thick edges in Fig.~\ref{fig:path}, and their colors show the overlap of the wavefunction decoded from the latent vectors and the actual DFT wavefunctions on the path.
The interpolated vectors exhibit a similar reconstruction overlap to those directly encoded latent vectors, thus demonstrating that the encoder preserves the smooth structure of the path.
The smoothness of the AE suggests its ability to infer the representations of previously unseen wavefunctions, and thus indicates the potential of such encoder-decoder-based representation for wavefunction generation.

The two $\Gamma$ points at the ends of the path do not coincide in the latent path of Fig.~\ref{fig:path} because they are related by an umklapp transformation.
The AE encodes the lattice-periodic part $u_{n\bm{k}}(\bm{r})$ of Bloch states rather than the full Bloch function $\psi_{n\bm{k}}(\bm{r})=e^{i\bm{k}\cdot\bm{r}}u_{n\bm{k}}(\bm{r})$.
This is because Bloch functions at neighboring $\bm{k}$ points are orthogonal, and encoding the full Bloch functions would therefore obscure the local relationship between neighboring points in the reciprocal space.
Under a reciprocal-lattice translation, or umklapp transformation, $\bm{k}\mapsto \bm{k}+\bm{G}$, where $\bm{G}$ is a reciprocal lattice vector, the full Bloch function remains invariant $\psi_{n\bm{k}+\bm{G}}(\bm{r})=\psi_{n\bm{k}}(\bm{r})$, but the lattice-periodic part undergoes a non-trivial linear transformation 
\begin{equation}
    u_{n\bm{k}+\bm{G}}(\bm{r})=e^{-i\bm{G}\cdot\bm{r}}u_{n\bm{k}}(\bm{r}).
\end{equation}
This relation requires a position-dependent phase rather than a global $U(1)$ phase, and thus $u_{n\bm{k}}(\bm{r})$ and $u_{n\bm{k}+\bm{G}}(\bm{r})$ are generally encoded to different latent vectors by the present AE.
Enforcing equivariance under reciprocal-lattice translations requires additional architectural constraints and is left for future work.
In practice, we can always restrict reciprocal space sampling to a fixed fundamental region, such as the first Brillouin zone or a parallelepiped reciprocal cell, and some additional boundary matching or periodicization procedures can be used if boundary continuity is important in the downstream application.

For comparison, we also study the action of gauge transformation in the real latent space produced by the non-$U(1)$-equivariant AE.
For a non-$U(1)$-equivariant real latent vector $\bm{z}=\mathrm{enc}(u)\in\mathbb{R}^{2d}$, the gauge transformation of the input, $u\mapsto ue^{i\theta}$, induces a general latent space map parameterized by phase angle $\theta\in[0,2\pi)$:
\begin{equation}
    \rho_{\theta}(\bm{z})=\mathrm{enc}(e^{i\theta}u).
\end{equation}
If the latent space transformed as a defining representation of $U(1)$, the gauge action would have the linear form 
\begin{equation}\label{eq:realU1}
    \rho_\theta(\bm{z})=\bm{z}\cos\theta+L\bm{z}\sin\theta,
\end{equation}
where $L$ is a linear operator with $L^2=-1$.

We find that, despite the absence of explicit equivariance constraints, the dependence of the actual gauge action on $\theta$ is well approximated by a sinusoidal form similar to Eq. \eqref{eq:realU1}:
\begin{equation}\label{eq:sine}
    \rho_\theta(\bm{z})\approx \bm{a}(\bm{z})\cos\theta+\bm{b}(\bm{z})\sin\theta - \bm{a}(\bm{z})+\bm{z},
\end{equation}
where the offset enforces $\rho_{\theta=0}(\bm{z})=\bm{z}$, i.e., the vector is itself without gauge transformation.
We sample $\theta$ at 32 values uniformly distributed in $[0,2\pi)$ and fit the sinusoidal form in Eq. \eqref{eq:sine} with respect to $\theta$ for 8,500 latent vectors of 128 dimensions encoded from the test dataset used in Fig.~\ref{fig:violin}.
The resulting fit has a median $r^2=0.9989$.
Fig. \ref{fig:realEquivariance}(a) shows the relative deviation between the directly computed $\rho_\theta(\bm{z})$ and the sinusoidal fit in $\theta\in[0,2\pi)$ for a representative latent vector $\bm{z}$ corresponding to the highest valence band of $\mathrm{MoS_2}$ at a randomly selected $\bm{k}$-point.
This example has $r^2=0.9988$, close to median quality.
The emergence of a sinusoidal structure with respect to $\theta$ may result from the arbitrary gauge from the eigensolver, which acts as an implicit data augmentation during training.
The residual shortcut structure in the non-$U(1)$ architecture may also contribute to this structure (see appendix \ref{sec:app-NN-AE}).

However, the sinusoidal dependence of $\rho_\theta(\bm{z})$ on $\theta$ does not imply that the non-equivariant latent space carries a linear representation of the $U(1)$ space, as the coefficients $\bm{a}(\bm{z})$ and $\bm{b}(\bm{z})$ can have nonlinear dependence on $\bm{z}$. 
Fig.~\ref{fig:realEquivariance}(b) shows the parity plot for the linear regressions of $\bm{a}(\bm{z})$ and $\bm{b}(\bm{z})$ against $\bm{z}$.
An affine fit gives a similar behavior.
While $\bm{a}(\bm{z})$ is approximately linear, with $r^2=0.996$ in the linear regression, $\bm{b}(\bm{z})$ exhibits stronger deviations from linearity, indicating that the gauge action $\rho_\theta(\bm{z})$ has a nontrivial dependence on the latent vector $\bm{z}$.
The asymmetry between $\bm{a}$ and $\bm{b}$ is likely due to the initial condition $\rho_{\theta=0}(\bm{z})=\bm{z}$.
Consequently, the gauge action in the non-equivariant latent space, $\rho_\theta(\bm{z})$, cannot be determined \textit{a priori} in a simple closed form, making it difficult to respect the gauge symmetry in downstream applications. 

\subsection{Topology}

\begin{figure*}
    \centering
    \includegraphics[width=0.99\textwidth]{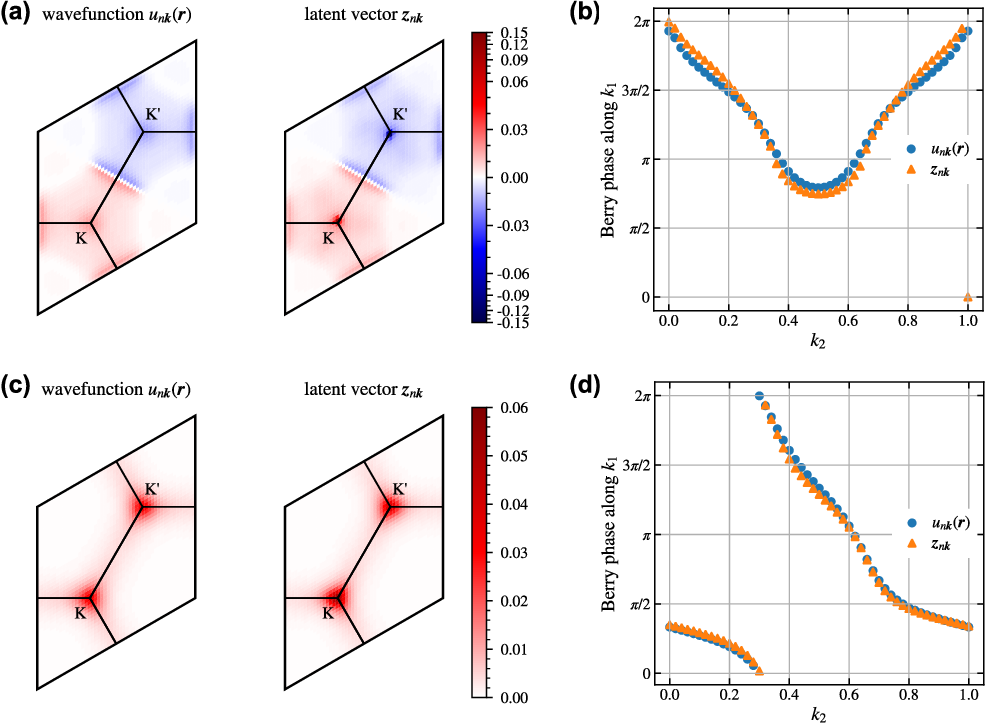}
    \caption{Topological properties of the wavefunctions $u_{n\bm{k}}$ and latent vectors $\bm{z}_{n\bm{k}}$.
    \textbf{(a)} Berry curvature of the $u_{n\bm{k}}$ and $z_{n\bm{k}}$ of the top valence band of monolayer $\text{MoS}_2$ in the parallelogram region. Linear scale is used for points with absolute values no greater than 0.06, while log scale is used for those above 0.06. 
    \textbf{(b)} Berry phase or wannier center winding of $u_{n\bm{k}}$ and $\bm{z}_{n\bm{k}}$ of the highest valence band of monolayer $\text{MoS}_2$.
    \textbf{(c)}\textbf{(d)} Berry curvature and Berry phase winding of a Haldane model built from graphene Wannier functions.
    }
    \label{fig:topology}
\end{figure*}

The gauge structure of wavefunctions across the Brillouin zone is closely related to band topology.
Due to $U(1)$ equivariance of the AE, we expect the latent vector to also contain information about the band topology.
Two quantities are used to examine this correspondence.
First, the Berry curvature \cite{Berry1984quantal, Xiao2010berry} at momentum $\bm{k}$ is the normalized Berry phase accumulated around an infinitesimal loop enclosing $\bm{k}$.
It quantifies the local geometric twisting of the Bloch states.
Second, the Brillouin-zone integral of Berry curvature yields the Chern number \cite{Thouless1982quantized}, which characterizes the global topological twisting of Bloch states of the band.
One way to identify the Chern number is through the winding of hybrid Wannier centers \cite{Gresch2017z2pack}. For a 2-dimensional reciprocal space with fractional coordinates ($k_1, k_2$), consider the loop obtained by traversing $k_1$ from 0 to 1 at a fixed $k_2$.
The Berry phase accumulated on the loop is a function of $k_2$, denoted by $\theta(k_2)$.
It is proportional to the hybrid Wannier center \cite{Gresch2017z2pack}.
The winding number of $\theta(k_2)$ modulo $2\pi$ as $k_2$ varies from 0 to 1 is equal to the Chern number of the band.
We calculate the Berry curvature and Berry phase winding using a discrete formalism \cite{Fukui2005chern,Vanderbilt2018berry} for both the wavefunctions $u_{n\bm{k}}(\bm{r})$ and latent vectors $\bm{z}_{n\bm{k}}$.
Details of the calculation are provided in appendix \ref{sec:app-topo-method}.

We find that the latent vectors inherit the single-band topological properties of the original wavefunctions.
Fig.~\ref{fig:topology}(a) shows the dimensionless discrete Berry curvature of the original wavefunction $u_{n\bm{k}}(\bm{r})$ and the corresponding 64-dimensional latent vectors $\bm{z}_{n\bm{k}}$ of the highest valence band of monolayer $\mathrm{MoS}_2$ in a parallelogram reciprocal cell.
The latent-space Berry curvature qualitatively resembles the curvature of the wavefunctions, reproducing features around the center and the edges of the $K / K'$ valleys, although quantitative differences remain, with the latent Berry curvature being more intense at the $K/K'$ valleys.
The quantitative difference between the Berry curvature of latent vectors and wavefunctions is provided in Fig. \ref{fig:topoDiff} of Appendix \ref{sec:app-topo-diff}.
The agreement between wavefunction and latent-space Berry curvature indicates that the latent vectors retain information about local geometric winding of Bloch states.
Fig. \ref{fig:topology}(b) shows the Berry phase winding of the wavefunction and latent vectors of the same band.
Both exhibit zero net winding, consistent with a zero Chern number.
As a further demonstration, we consider a topologically nontrivial Chern band of a Haldane model \cite{haldane1988model} constructed from graphene Wannier functions, as shown in Fig.~\ref{fig:topology}(c)(d).
Additional details of constructing the haldane model are provided in appendix \ref{sec:app-topo-haldane}.
A Chern number 1 can be identified from the winding of both the wavefunction and the latent vector.
These results indicate that the equivariant latent representation captures information of both local twisting structure in Berry curvature and the global twisting topology in a single band.

A limitation to the topology of latent vectors arises from the lack of exact umklapp relations, as discussed in section \ref{sec:equivariance}.
In Fig. \ref{fig:topology}, the wavefunctions at the boundaries $k_1=1$ and $k_2=1$ are explicitly obtained from a reciprocal lattice translation of the wavefunctions at $k_1=0$ and $k_2=0$, and the latent vectors at these boundaries are directly encoded from corresponding wavefunctions.
Because exact umklapp equivariance is not enforced, the latent vectors on opposite boundaries do not have an exact linear relation.
Consequently, the Berry curvature of the latent vectors is only approximately periodic with respect to the reciprocal lattice vectors, and the latent space winding number in Fig. \ref{fig:topology}(c)(d) is close to, but not exactly by construction, an integer.
Nevertheless, the latent vectors contain sufficient information to correctly identify the band Chern number as an integer, and to give a qualitative description of the band Berry curvature.
Indeed, summing the discrete latent Berry curvature and dividing by $2\pi$ yields -0.0068 for $\text{MoS}_2$ and 1.0018 for the Haldane model, with deviations of only $10^{-3}$-$10^{-2}$ from the exact values of 0 and 1.

\begin{figure}[!t]
    \centering
    \includegraphics[width=0.9\linewidth]{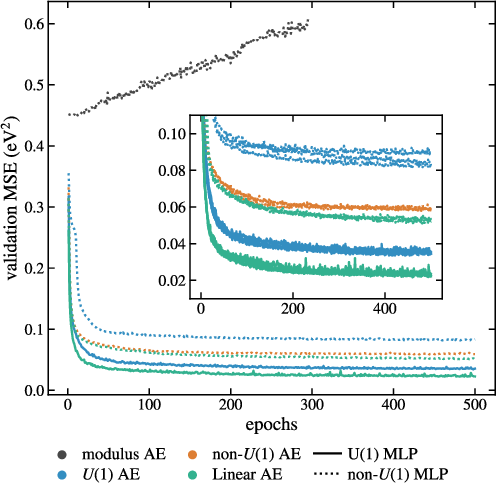}
    \caption{Training Curve of different latent vectors and multilayer perceptrons (MLPs) on the validation set. Latent vectors from different autoencoders (AEs) are shown in different colors, and different MLPs are marked as different linestyles.
    The main panel is a comparison of the AE encoding only the modulus of wavefunctions (black) and other AEs also encoding the phase information, including the $U(1)$ AE (blue), the non-$U(1)$ AE (orange) and the linear AE (green).
    The inset is a zoomed-in comparison of different AEs retaining phase information combined with either the $U(1)$ MLP (solid lines) or non-$U(1)$ MLP (dotted lines).
    The versions with phase information are run with three different random seeds.
    }
    \label{fig:GWMatTrain}
\end{figure}

The similarity between the topological properties derived from $\bm{z}_{n\bm{k}}$ and those from $u_{n\bm{k}}(\bm{r})$ is a direct consequence of $U(1)$ equivariance and the local smoothness of the AE.
Consider the normalized overlap of neighboring wavefunctions
\begin{equation}\label{eq:link}
    M_\mu(\bm{k})=\frac{\langle u_{\bm{k}}|u_{\bm{k}+\bm{\hat{\mu}}\Delta k }\rangle}{|\langle u_{\bm{k}}|u_{\bm{k}+\bm{\hat{\mu}}\Delta k }\rangle|}\equiv \exp\big(i\theta_{\mu}(\bm{k})\big),
\end{equation}
where $\mu$ is a direction in the Brillouin zone and $\Delta k$ is some small displacement.
The band index is suppressed here because only a single band is considered.
The quantity $M_\mu(\bm{k})$ is directly related to the Berry connection \cite{Vanderbilt2018berry} and is a fundamental quantity in the discrete computation of topological properties.
For sufficiently small $|\Delta k|$, the overlap of neighboring wavefunctions is close to 1, i.e., $|\langle u_{\bm{k}}|u_{\bm{k}+\bm{\hat{\mu}}\Delta k }\rangle|\approx 1$.
Thus, after removing the relative phase, $\langle u_{\bm{k}}|e^{-i\theta_\mu(\bm{k})}u_{\bm{k}+\hat{\bm{\mu}}\Delta k}\rangle\approx 1$, implying that the numerical values of the two wavefunctions are close, so that $u_{\bm{k}}(\bm{r})\approx e^{-i\theta_\mu(\bm{k})}u_{\bm{k}+\hat{\mu}\Delta k}(\bm{r})$.
If the encoder ($\mathrm{enc}$) is smooth, these nearby inputs are mapped to nearby latent vectors.
We therefore write
\begin{equation}
    \frac{\langle \mathrm{enc}(u_{\bm{k}})|\mathrm{enc}(e^{-i\theta_\mu(\bm{k})}u_{\bm{k}+\hat{\bm{\mu}}\Delta k})\rangle}{|\langle \mathrm{enc}(u_{\bm{k}})|\mathrm{enc}(e^{-i\theta_\mu(\bm{k})}u_{\bm{k}+\hat{\bm{\mu}}\Delta k})\rangle|}=e^{i\delta(\bm{k})},
\end{equation}
where $\delta(\bm{k})$ denotes the phase error introduced by encoding.
Using the $U(1)$ equivariance of the encoder, the link variable of latent vectors $\bm{z}_{\bm{k}}=\mathrm{enc}(u_{\bm{k}})$ and $\bm{z}_{\bm{k}+\hat{\bm{\mu}}\Delta k}=\mathrm{enc}(u_{\bm{k}+\hat{\bm{\mu}}\Delta k})$ becomes
\begin{equation}
    \tilde{M}_\mu(\bm{k})=\frac{\langle z_{\bm{k}}|z_{\bm{k}+\bm{\hat{\mu}}\Delta k }\rangle}{|\langle z_{\bm{k}}|z_{\bm{k}+\bm{\hat{\mu}}\Delta k }\rangle|}\equiv \exp\big[i\big(\theta_{\mu}(\bm{k})+\delta(\bm{k})\big)\big].
\end{equation}
When the encoder is smooth enough such that $\delta(\bm{k})$ is small, wavefunctions and latent vectors have similar link variables $M_\mu(\bm{k})\approx\tilde{M}_\mu(\bm{k})$ and thus similar topological properties.
The larger discrepancies near the $K/K'$ valleys (Appendix Fig. \ref{fig:topoDiff}(a)) suggest that the encoder is less smooth for wavefunctions around these regions.
The exact cause of this behavior is related to dataset sampling and training dynamics, and is difficult to analyze exactly.
In addition, the Berry curvature may have a larger relative deviation than the Berry phase, as also observed in Fig.~\ref{fig:topology}.
This is because the Berry curvature corresponds to a derivative-like quantity of the link variable, and the difference of the phase $\theta(\bm{k})$ at neighboring points can be much smaller than the individual link phase.
Small phase errors in the encoding can thus produce large relative errors in the local Berry curvature.

\section{Prediction using Latent Representations}\label{sec:prediction}

As a proof-of-principle demonstration of the importance of phase information and $U(1)$-equivariance in downstream prediction using the latent representation, we train a simple multilayer perceptron (MLP) to predict the off-diagonal elements of the $G_0W_0$ Generalized Plasmon-Pole (GPP) model self-energy matrix \cite{Hybertsen1986electron} of two-dimensional materials:
\begin{equation}\label{eq:offdiag-sigma}
    \Sigma_{mn}(\bm{k})\equiv\left \langle m\bm{k}\left |\frac{\Sigma(E^{DFT}_{m\bm{k}})+\Sigma(E^{DFT}_{n\bm{k}})}{2}\right|n\bm{k} \right \rangle,
\end{equation}
where $E_{m\bm{k}}^{DFT}$ and $E_{n\bm{k}}^{DFT}$ are the DFT energies of state $|m\bm{k}\rangle$ and $|n\bm{k}\rangle$.
The structure in Eq. \eqref{eq:offdiag-sigma} makes the resulting operator Hermitian, which is important in quasiparticle self-consistent GW theory \cite{vanSchilfgaarde2006quasiparticle, Kotani2007quasiparticle}.
Additional details about generating the GW data are provided in appendix \ref{sec:app-data}.

A $U(1)$-equivariant MLP is constructed to respect the gauge equivariance of the off-diagonal matrix elements.
A schematic of the architecture is shown in Fig.~\ref{fig:architecture}(c).
The model takes two latent vectors $\bm{z}_{m\bm{k}}$ and $\bm{z}_{n\bm{k}}$, and constructs an equivariant coupled feature vector through the coupling layer
\begin{equation}
    \bm{f}_{mn\bm{k}}=\sum_{p=1}^{P} A_p \bm{z}_{m\bm{k}}^*\,\odot\,B_p\bm{z}_{n\bm{k}},
\end{equation}
where $A_p$ and $B_p$ are trainable weight matrices and $\odot$ denotes elementwise multiplication.
The coupled feature $\bm{f}_{mn\bm{k}}$ subsequently goes through an ordinary MLP with complex weights and equivariant nonlinearities \eqref{eq:U1activation}.
Such an architecture guarantees the correct gauge equivariance of off-diagonal matrix elements of $\Sigma_{mn}(\bm{k})$.
In addition, the weight matrices in the MLP are conditioned on additional scalar inputs consisting of the DFT energies, the momentum, and lattice structure including the two in-plane lattice constants and the angle of the two lattice basis vectors.
For comparison, we also construct a non-$U(1)$ equivariant MLP where all complex-valued layers are replaced with real-valued counterparts, and equivariant nonlinearities are replaced with conventional nonlinearities.
Additional details of the MLP architecture are provided in appendix \ref{sec:app-NN-mlp}.
We emphasize that this simple 2-state MLP is not a quantitatively complete model able to accurately predict the self-energy matrix $\Sigma_{mn}(\bm{k})$, as $\Sigma_{mn}(\bm{k})$ in principle depends on the full set of mean-field states $\{|n\bm{k}\rangle,E_{n\bm{k}}\}$, and an accurate prediction model should thus accept all relevant states as input.
The present MLP architecture is used as a proof-of-principle model to isolate the roles of phase information and gauge equivariance.

Fig.~\ref{fig:GWMatTrain} shows the training curve of the MLPs on the validation set for different latent vectors and MLPs.
The latent vectors produced by 4 different AEs are compared and depicted by different colors: an AE encoding only the modulus $|u_{n\bm{k}}|$ (modulus AE, black), a nonlinear $U(1)$-equivariant AE ($U(1)$ AE, blue), a nonlinear non-$U(1)$ equivariant AE (non-$U(1)$ AE, orange), and a linear AE (linear AE, green).
The equivariant and non-equivariant MLPs are compared for these latent vectors, indicated by solid and dotted lines, respectively.
The MLPs are trained on a dataset consisting of 613,200 matrix elements from 438 2-dimensional materials with hexagonal symmetry, with a train-validation split of $0.9:0.1$ across the randomly shuffled matrix elements.
The dataset is reduced compared with the wavefunction dataset in section \ref{sec:u1ae} due to the high computational cost of generating self energy matrix data.
For controlled comparison, the sizes of models are kept as approximately the same.
All AEs use approximately 5.0M real parameters and 64 complex dimensions or 128 real dimensions, including the modulus-only AE.
All MLPs use approximately 155k real parameters.

The results demonstrate that both phase information and gauge equivariance are important to the prediction of gauge-sensitive quantities such as $\Sigma_{mn}(\bm{k})$. Though all model variants have some level of overfitting (see appendix \ref{sec:app-mlp}), likely due to the non-completeness of inputs, there are clear distinctions between different representations and prediction architectures.
The main panel of Fig.~\ref{fig:GWMatTrain} compares modulus-only representation and other representations with phase information.
The validation loss of the modulus-only version increases as training proceeds, and is much larger than those of the other models, suggesting severe overfitting.
The inset of Fig.~\ref{fig:GWMatTrain} is a zoomed-in comparison among the versions retaining phase information.
Models that neglect $U(1)$ equivariance in either the AE or the downstream MLP perform worse than the fully equivariant models.
It indicates that respecting gauge equivariance in both the representation learning and the prediction stage improves the learning of gauge-sensitive matrix elements.

Two observations in Fig.~\ref{fig:GWMatTrain} differ from our initial expectations.
First, the linear AE yields better downstream MLP prediction performance than the full nonlinear $U(1)$-equivariant AE, despite having significantly worse reconstruction quality, as shown in Fig.~\ref{fig:violin}.
One possible explanation is that the nonlinear AE encodes the wavefunction information into a more complicated latent space that cannot be effectively decoded by the simple MLP used here.
Second, the combination of the $U(1)$-equivariant AE with the non-equivariant MLP performs worse than the fully non-equivariant combination.
A possible reason is an architectural mismatch: the $U(1)$-equivariant AE fully respects the gauge and complex structure of wavefunctions, which is completely ignored in the non-$U(1)$ MLP, and thus the downstream non-equivariant MLP may fail to utilize the information in the equivariant latent representation.
These two observations both suggest that representation learning and downstream prediction architectures should be designed jointly in future studies.

\section{Discussion}\label{sec:discussion}

Gauge equivariance has been previously investigated in several areas of machine learning.
It has been studied in the context of local coordinate transformations \cite{Cohen2019gauge, Haan2020gauge}, and has been widely utilized in machine learning methods for lattice gauge theories, including normalizing flows \cite{Kanwar2020equivariant}, neural-network quantum states \cite{Luo2021gauge} and general predictive models \cite{Favoni2022lattice}.
In condensed matter applications, a gauge-equivariant neural network has also been used to predict multiband Chern numbers from the Wilson loop \cite{Huang2025learning}.
Here, we address a distinct problem: the representation learning of individual Bloch states, where the arbitrary $U(1)$ phase of each state constitutes a redundancy that must be handled if the full complex wavefunction is to be retained. Our results show that this gauge structure can be built directly into a compact latent representation and subsequently propagated into gauge-sensitive downstream models.

Intriguingly, we find that the non-$U(1)$-equivariant AE reconstructs the wavefunctions as accurately as the equivariant AE, and also develops an approximate gauge structure in the non-equivariant latent space.
One possible explanation is that the gauge symmetry reduces to a global $U(1)$ symmetry in the AE, which may be easier for the model to learn from the implicit data augmentation induced by the random gauge adopted by the eigen solver. This result highlights an important distinction between reconstruction accuracy and the transformation properties of the learned representation. Although explicit $U(1)$ equivariance is not required for reconstruction, enforcing it gives the latent vectors an exact and known gauge transformation law. This structure can then be directly incorporated into downstream models of gauge-sensitive quantities, as demonstrated by the self-energy matrix element prediction. The equivariant representation additionally retains geometric information about the underlying Bloch states, allowing for the calculation of Berry curvature and Chern number directly from the latent vectors.

We note that the downstream prediction considered here is intended as a proof-of-principle test of whether the learned representations can be used to predict a quantity that transforms nontrivially under the gauge freedom of the input Bloch states.
The deliberately restricted model therefore isolates the role of phase information and the correct $U(1)$ transformation law, rather than serving as a quantitative surrogate for $GW$.
Extending this construction to a many-state architecture would be required for quantitative prediction of the full self-energy matrix.

The present study encodes each Bloch state independently.
Explicitly incorporating relationships between different crystal momenta and different bands is  an important direction for future studies.
First, spatial symmetry is not incorporated in our current architecture.
It could be incorporated by combining $U(1)$ equivariance with space-group equivariance, so that wavefunctions at symmetry-equivalent points can be mapped to related latent vectors.
Second, as is discussed in section \ref{sec:equivariance}, umklapp relations are not explicitly enforced in our present approach, so latent vectors at $\bm{k}$ and $\bm{k}+\bm{G}$ do not have explicit relations by construction.
Third, when degeneracy is important, the relevant gauge freedom is generally a multiband $U(N)$ symmetry for $N$ bands, rather than $U(1)$.
Extending the current representation-learning framework to enforce umklapp equivariance and multiband $U(N)$ symmetry will require additional neural network architectural structure.

To conclude, we introduce gauge equivariance into machine learning models that use electronic features learned from individual Bloch states.
We show that $U(1)$ equivariance endows the latent representation with physically meaningful properties including an exact and explicit gauge transformation law, smooth variation in reciprocal space, and information about band topology.
Importantly, these properties are obtained without a significant change in reconstruction accuracy relative to a non-equivariant autoencoder, demonstrating that reconstruction fidelity alone does not determine the physical structure of the latent representation.
We further demonstrate that retaining the phase information of the wavefunctions and respecting gauge equivariance are important in the prediction of gauge-sensitive objects, as illustrated by the prediction of off-diagonal $GW$ self-energy matrix elements.
This work provides a route for incorporating phase-sensitive electronic-structure information contained in electronic states into machine-learning representations while preserving their underlying gauge structure, enabling their use in models of operators, geometric quantities, and many-body properties beyond scalar observables.
\section{Acknowledgment}

This work was primarily supported by the U.S. Department of Energy, Office of Science, Basic Energy Sciences Condensed Matter Theory Program  Award No. DE-SC0026655. Development of the BerkeleyGW code was supported by Center for Computational Study of Excited-State Phenomena in Energy Materials (C2SEPEM) at the Lawrence Berkeley National Laboratory, funded by the U.S. Department of Energy, Office of Science, Basic Energy Sciences, Materials Sciences and Engineering Division, under Contract No. DE-C02-05CH11231.
The calculations used resources of the National Energy Research Scientific Computing (NERSC), a DOE Office of Science User Facility operated under contract no. DE-AC02-05CH11231 under award numbers BES-ERCAP-0031507 and BES-ERCAP-0027380. An award of computer time was provided by the U.S. Department of Energy’s (DOE) Innovative and Novel Computational Impact on Theory and Experiment (INCITE) Program. This research used supporting resources at the Argonne and the Oak Ridge Leadership Computing Facilities. The Argonne Leadership Computing Facility at Argonne National Laboratory is supported by the Office of Science of the U.S. DOE under Contract No. DE-AC02-06CH11357 . The Oak Ridge Leadership Computing Facility at the Oak Ridge National Laboratory is supported by the Office of Science of the U.S. DOE under Contract No. DE-AC05-00OR22725.

\appendix

\section{Details of Neural Network Architecture and Training} \label{sec:app-NN} 
\subsection{Autoencoder} \label{sec:app-NN-AE}

\begin{table*}[!htbp]
    \caption{Hyperparameters of the $U(1)$ Autoencoder \label{stab:hyperparamAE}}
        \begin{tabular}{c|c}
        \hline\hline
         Hyperparameter    & values \\
         \hline
         Number of compression / upsampling layers & 5\\
         Output channel dimensions of each compression layer & 120, 144, 144, 96, 48 \\
         Stride of each pooling layer (stride 1 means no pooling) & 2, 2, 2, 1, 1 \\
         Kernel size of each compression layer & 3, 3, 3, 3, 3\\
         Group size in group norm & 8\\
         hidden dimension of MLP in nonlinearity & 16\\
         Denominator regularization $\epsilon$ & $10^{-5}$\\
         \hline\hline
        \end{tabular}
\end{table*}

We use 2-dimensional complex convolution as building blocks of the 2-dimensional autoencoder (AE) architecture.
Let $\bm{x}\in\mathbb{C}^{C\times H\times W}$ be the input tensor, with components $x_{cij}$, where $c=0,\dots,C-1$ is the channel dimension, and $i=0,\dots,H-1$, $j=0,\dots,W-1$ are the spatial dimensions.
Let $\mathrm{conv}$ be a convolution with output dimension $C'$, padding $p$ on all four edges, kernel size $k$ and stride $s$.
The output $\mathrm{conv}(\bm{x})$ is in $\mathbb{C}^{C'\times H' \times W'}$, where
\begin{equation}
    H'=\left\lfloor \frac{H+2p-k}{s} +1\right\rfloor,\quad W'=\left\lfloor \frac{W+2p-k}{s} +1\right\rfloor.
\end{equation}
The components of the outputs are
\begin{equation}
    \mathrm{conv}(\bm{x})_{c'i'j'}=\sum_{c=0}^{C-1} \sum_{m=0}^{k-1} \sum_{n=0}^{k-1} K_{c'c\,mn}\,x_{c,\,i's+m-p,\,j's+n-p}
\end{equation}
where $K\in\mathbb{C}^{C'\times C\times k\times k}$ is the trainable kernel.

The autoencoder consists of an encoder and a decoder. As input to the encoder, the wavefunctions of 2-dimensional materials are represented on a grid as $\bm{u}\in \mathbb{C}^{N_3\times (N_1\times N_2)}$ with components $u_{kij}$, for $N_{1},N_2,N_3$ grid points along the three crystal axes.
The input is normalized so that each pixel is on the order of 1:
\begin{equation}\label{seq:norm}
\frac{1}{N_1N_2N_3}\sum_{kij}^{N_1N_2N_3}|u_{kij}|^2=1
\end{equation}
The out-of-plane direction $N_3$ is treated as a channel index.
The input $\bm{u}$ first undergoes $L$ compression blocks and is transformed to $\bm{u}^{(L)}\in\mathbb{C}^{C\times N_1'\times N_2'}$ with smaller spatial dimension $N_1'$ and $N_2'$.
Each compression block contains a residual convolution (Fig. 1(b)) and an anti-aliased pooling.
The residual convolution changes the channel dimension while keeping the spatial dimension.
All convolutions with kernel size $k$ are chosen to have stride 1 and padding $(k-1)/2$.
Circular padding is used to respect the periodicity of the wavefunctions.
Complex group norm \cite{Wu2018Group} is used in the residual block.
To achieve $U(1)$ equivariance, the affine transform in the group norm is calculated as
\begin{equation}
    \tilde{\bm{x}}= \gamma \bm{x}+b\frac{\mu}{|\mu|+\epsilon},
\end{equation}
where $\mu$ is the mean of the group, $\epsilon$ is a small positive number for regularization and $\gamma\in\mathbb{R}$, $b\in\mathbb{C}$ are trainable parameters.
The nonlinearity used in the residual block is an element-wise complex function
\begin{equation}
    a(x)=\mathrm{MLP}(|x|)\frac{x}{|x|+\epsilon},\quad x\in\mathbb{C}
\end{equation}
where $\mathrm{MLP}$ is a real 2-layer small multilayer perceptron with softplus activation.
The pooling layer is also a convolution.
Its kernel is trivial with respect to the channel dimension $K_{c'cmn}\propto\delta_{cc'}$, and is a Gaussian distribution along the spatial dimension to suppress high-frequency signals.
A stride greater than 1 is used in pooling layers for downsampling.
After the compression layers, a dense linear layer is used to compress $\bm{u}^{(L)}\in\mathbb{C}^{C\times N_1'\times N_2'}$ to $\mathbb{C}^d$, where $d$ is the latent dimension, after which the latent vector is normalized as
\begin{equation}
    \frac{1}{d}\sum_{i=1}^d|z_i|^2=1.
\end{equation}
In the decoder, a dense linear layer is first used to map $\mathbb{C}^d$ to $\mathbb{C}^{C\times N_1'\times N_2'}$ so that the shape $\bm{u}^{(L)}$ before the dense linear compression is restored.
After that, $L$ upsampling layers consisting of bilinear upsampling layers and residual convolution blocks are used to reconstruct the wavefunctions.
The evolution of channel and spatial dimension in the decoder exactly mirrors that in the encoder.
In the end, the reconstructed wavefunction is normalized according to Eq. \ref{seq:norm}.
The hyperparameters used in the AE are listed in Table \ref{stab:hyperparamAE}.

The architecture of the non-$U(1)$-equivariant AE is the same as the equivariant AE, except the following modifications: First, the convolutions and group norms are replaced with real-valued layers, and bias is used in the convolutions.
Second, the GELU nonlinearity is used to replace the equivariant nonlinearities.
Third, the nonlinearity at the output of the residual block after the shortcut connection is discarded, so that the residual block contains only one nonlinearity.
The last modification is because real ReLU-like activations discard half of the information, and we find that discarding the second nonlinearity improves the reconstruction of the non-equivariant version to be similar to the equivariant version.
Note that the equivariant activation does not have this problem.
In the non-equivariant autoencoder, the channel dimensions are chosen as 176, 200, 200, 128, 64 so that the total number of parameters approximates the equivariant version.

By deleting the second activation in the residual block, the whole non-equivariant encoder can be written as a globally linear part $W$ plus a nonlinear part $f$:
\begin{equation}
    \bm{z}=\mathrm{enc}(\bm{u})=W\bm{u}+f(\bm{u}),
\end{equation}
where the linear part comes from all the shortcut connections in the residual blocks.
In the non-equivariant AE, the input wavefunction is taken as the concatenation of the real part $\bm{u}_r$ and the imaginary part $\bm{u}_i$: $\bm{u}=[\bm{u}_r, \bm{u}_i]$.
Under a $U(1)$ transformation, the input wavefunction transforms as
\begin{equation}
    \bm{u}\mapsto [\bm{u}_r\cos\theta-\bm{u}_i\sin\theta,\,\bm{u}_r\sin\theta+\bm{u}_i\cos\theta].
\end{equation}
If the encoder is exactly linear, each component of the latent vector will transform in a sinusoidal form $a\cos\theta+b\sin\theta$ under the $U(1)$ gauge transformation, and $a$ and $b$ are both linearly dependent on $\bm{z}$.
This fact may partly contribute to the approximate sinusoidal form of gauge action in the non-equivariant latent space.

When training the autoencoders, a learning rate of $2\times 10^{-4}$ is used.
Learning rate will be halved if no improvement occurs on the validation set in 40 epochs. 
The training is stopped if no improvement occurs on the validation set in 60 epochs or 1200 epochs is reached.
The weights with the smallest error on validation set are saved as the final result.

\subsection{Multilayer Perceptron} \label{sec:app-NN-mlp}
The multilayer perceptron (MLP) used to predict self energy matrices consists of a coupling layer and a $U(1)$-equivariant MLP.
To predict a target matrix element $\Sigma_{mn\bm{k}}=\langle m\bm{k}|\Sigma|n\bm{k}\rangle$, the two $d$-dimensional latent vectors $\bm{z}_{m\bm{k}}$ and $\bm{z}_{n\bm{k}}$, corresponding to the basis states are taken as input.
First, a coupled feature vector is computed through the coupling layer
\begin{equation}
    \bm{f}_{mn\bm{k}}=\sum_{p=1}^P\, A_p\bm{z}_{m\bm{k}}^* \odot B_p\bm{z}_{n\bm{k}},
\end{equation}
where $\odot$ is elementwise multiplication, $A_p,B_p$ are trainable matrices, and $\bm{f}_{mn\bm{k}}$ is the coupled feature vector.
Under a gauge transformation, the coupled feature $\bm{f}_{mn\bm{k}}$ transforms in the same way as the matrix element $\Sigma_{mn\bm{k}}$.
The coupled feature then undergoes a $U(1)$-equivariant MLP consisting of linear trainable matrices an element-wise activations chosen as
\begin{equation}
    a(x)=\mathrm{LeakyReLU}(|x|-b)\frac{x}{|x|+\epsilon},\quad x\in\mathbb{C},
\end{equation}
where $b$ is a learnable parameter.
Finally, the feature vector output by the last hidden layer is linearly mapped to a complex scalar to approximate the matrix element $\Sigma_{mn\bm{k}}$.
In the MLP, all linear weights are conditioned on scalar features including the DFT energies $E_{m\bm{k}}$ and $E_{n\bm{k}}$, the momentum $\bm{k}$, and lattice information including the lattice parameters $a_{1,2}$ along two in-plane crystal axes and the angle of the two in-plane basis vectors $\cos\alpha$, assembled as $\bm{v}=(E_{m\bm{k}}^{DFT}, E_{n\bm{k}}^{DFT}, \bm{k},a_1, a_2,\cos\alpha)\in\mathbb{R}^{d_v}$.
Each weight matrix $W$, including those in the coupling layer and the MLP, is computed as
\begin{equation}
    W(\bm{v})=W_0+\sum_{q=1}^Q (L_q\bm{v})(R_q\bm{v}),
\end{equation}
where $W_0$, $L_q$ and $R_q$ are trainable weights.
The total weight is $W(\bm{v})\in\mathbb{C}^{d_1\times d_2}$, and the condition matrices $L_q\in\mathbb{C}^{(d_1\times r)\times d_v}$, $R_q\in\mathbb{C}^{(r\times d_2)\times d_v}$ have lower rank $r<\mathrm{min}(d_1,d_2)$. The hyperparameters of the prediction multilayer perceptron are listed in Table \ref{stab:hyperparamMLP}.

\begin{table}[!htbp]
    \caption{Hyperparameters of the $U(1)$ MLP \label{stab:hyperparamMLP}}
        \begin{tabular}{c|c}
        \hline\hline
         Hyperparameter    & values \\
         \hline
         Dimension of combined feature & 48\\
         Hidden dimensions in MLP & 32,24,18\\
         Number of matrices $P$ in the coupling layer & 6\\
         Number of matrices $Q$ in condition layers & 2\\
         Rank $r$ of condition matrices & 8 \\
         \hline\hline
        \end{tabular}
\end{table}

The architecture of the non-equivariant MLP is the same as the equivariant version, except that all the layers are replaced with real-valued ones, and the activation is replaced by plain LeakyReLU.
The coupled feature dimension is chosen as $60$ and the hidden dimensions are chosen as 45,30,20 in the non-equivariant MLP, so that the number of parameters is similar to the equivariant version.

When training the prediction MLPs, a learning rate of $3\times 10^{-4}$ is used, and  a weight decay of $10^{-4}$ is used to reduce overfitting. All MLPs are trained for 500 epochs, and the weights with the smallest validation error are saved as the final result.

\section{Dataset Preparation} \label{sec:app-data} 
The materials studied in this work all come from the Computational 2D Materials Database (C2DB) \cite{Haastrup2018computational, Gjerding2021recent}.
We keep only the monolayers with a band gap between $0.5$ eV and $5$ eV and with an in-plane cell area smaller than $42\,\text{\AA}^2$. 1638 structures comprising 62 different elements are obtained in this way.
After that, similar to \cite{Hou2025mbformer}, structures with no more than 6 atoms in the cell are chosen, leading to 834 materials in total, which are used in the training and testing of the autoencoders.
Here, the lower bound of the gap is used to exclude metals, and the bounds of the cell area and number of atoms is mainly for considerations of computational cost.
The out-of-plane lattice constant of each structure is fixed to $12\,\text{\AA}$, hence all the materials have the same cell length along the out-of-plane direction.

The mean-field wavefunctions are calculated using the Quantum ESPRESSO software package\cite{Giannozzi2009quantum, Giannozzi2017advanced}, where the Perdew-Burke-Ernzerhof (PBE) exchange-correlation functional \cite{perdew1996generalized} and the scalar-relativistic optimized norm-conserving Vanderbilt (ONCV) pseudopotentials \cite{hamann2013optimized} are adopted.
The plane-wave cutoff for the wavefunctions is chosen to be $50$ Ry, and a $\Gamma$-centered $5\times5\times1$ grid covering the whole Brillouin zone is used for the $\bm{k}$-point sampling.
Spin-orbit coupling is not considered.
We also turn off all the crystal symmetries, so that the wavefunction at every $\bm{k}$ point is obtained from a direct diagonalization instead of from symmetry unfolding.
Four valence bands and six conduction bands are used to train the autoencoders.
During training, the wavefunctions are represented on a $24\times24\times 40$ real-space grid, which retains an average charge of 0.9994.

The self-energy matrices are calculated by BerkeleyGW \cite{Deslippe2012berkeleygw} in the one-shot $G_0W_0$ approximation, and the frequency dependence of the screening is described by the generalized plasmon-pole model \cite{Hybertsen1986electron}.
The inverse dielectric matrix is computed on the same $5\times5\times1$ grid with a dielectric cutoff of $25$ Ry.
Coulomb truncation \cite{IsmailBeigi2006truncation} is applied along the out-of-plane direction to remove the interaction between the periodic images of the monolayer.
Since the screening of a 2-dimensional material changes rapidly near $\bm{q}=0$, the nonuniform neck subsampling scheme with 10 extra $\bm{q}$ points \cite{Jornada2017nonuniform} is further used in this region.
The empty states entering the dielectric matrix and the Coulomb-hole summation are generated in the full plane-wave basis, and are then compressed to about 250 effective bands for each material by the stochastic pseudobands technique \cite{Altman2024mixed}.
Finally, the self-energy is evaluated at all 25 $\bm{k}$ points, and for each $\bm{k}$ point the 4 highest valence bands and the 4 lowest conduction bands are included.

\section{Calculation of Topological Properties} \label{sec:app-topo}

\subsection{Calculation Methods}\label{sec:app-topo-method}

Consider a single band of vectors dependent on crystal momentum, denoted as $|u(\bm{k})\rangle$.
In a 2D reciprocal space, take a grid in fractional coordinates
\begin{equation}
\bm{k}^l=\frac{j_1}{N_1}\bm{g}_1+\frac{j_2}{N_2}\bm{g}_2,
\end{equation}
with $\bm{g}_{1,2}$ being the two basis vectors of the reciprocal lattice, $j_\mu=0,1,2,\dots,N_\mu$ for $\mu=1,2$ and $l=(j_1,j_2)$.
The vectors at the edge $j_\mu=N_\mu$ should be taken as the umklapp of the vector at $j_\mu=0$.
The discrete Berry curvature is calculated as \cite{Fukui2005chern}:
\begin{equation}
    F(\bm{k}_l)=\mathrm{arg}[M_1(\bm{k}^l)\,M_2(\bm{k}^l+\hat{1})M_1(\bm{k}^l+\hat{2})^{-1}M_2(\bm{k}^l)^{-1}],
\end{equation}
where $\mathrm{arg}$ denotes the phase angle of a complex number in $(-\pi,\pi]$ and 
\begin{equation}
    M_\mu(\bm{k}^l)=\frac{\big\langle u(\bm{k}^l)|u(\bm{k}^l+\hat{\mu})\big\rangle}{\big|\big\langle u(\bm{k}^l)|u(\bm{k}^l+\hat{\mu})\big\rangle\big|},
\end{equation}
denotes the link variable, with $\mu=1,2$ being two directions of the reciprocal space, and $\bm{k}^l+\hat{\mu}\equiv \bm{k}^l+\bm{g}_\mu/N_\mu$ being the next neighboring grid point to $\bm{k}^l$ along the $\hat{\mu}$ direction.
The summation of the Berry curvature on the grid divided by $2\pi$ is the Chern number, which is guaranteed to be an integer in the case of exact periodicity or umklapp relation \cite{Fukui2005chern}.
We also identify the Chern number via hybrid Wannier center winding \cite{Gresch2017z2pack}.
In the discrete formulation, take a line along direction $1$ at a certain $k_2$ with $N_1+1$ points on the line:
\begin{equation}
    \bm{k}^i=\frac{i}{N_1}\bm{g}_1+k_2\,\bm{g}_2,
\end{equation}
with $k_2=j_2/N_2$ for $j_2=0,1,2,\dots,N_2$.
The product of the link variables along the line is the Wilson loop:
\begin{equation}
    W(k_2)=\prod_{i=0}^{N_1-1}\frac{\big\langle u(\bm{k}^i)|u(\bm{k}^{i+1})\big\rangle}{\big|\big\langle u(\bm{k}^i)|u(\bm{k}^{i+1})\big\rangle\big|},
\end{equation}
which is a function of $k_2$.
Its phase angle is a Berry phase \cite{Vanderbilt2018berry} dependent on $k_2$:
\begin{equation}
    \theta(k_2)=-i\ln W(k_2)\mod 2\pi, 
\end{equation}
with the physical meaning of a Wannier center, and the winding number of $\theta(k_2)$ as $k_2$ goes from 0 to 1 equals the Chern number \cite{Gresch2017z2pack}.


\subsection{Quantitative Difference between Latent Vectors and Wavefunctions} \label{sec:app-topo-diff}

\begin{figure}[!htbp]
    \centering
    \includegraphics[width=0.9\linewidth]{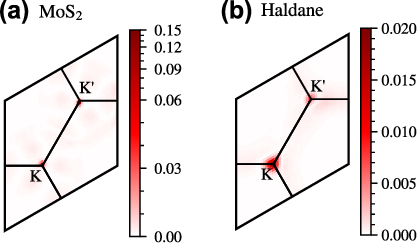}
    \caption{Difference between the wavefunction Berry curvature $F_{\text{wfn}}$ and latent vector Berry curvature $F_{\text{latent}}$, $|F_{\text{wfn}}-F_{\text{latent}}|$, for $\text{MoS}_2$ top valence band \textbf{(a)} and the Haldane model \textbf{(b)}. In panel (a), linear scale is used for values below 0.06, while log scale is used for those above 0.06.}
    \label{fig:topoDiff}
\end{figure}

We use the discrete method above to calculate topological properties of both the wavefunctions and the latent vectors. Fig. \ref{fig:topoDiff} shows the difference between the Berry curvature of the original wavefunction and the latent vector for $\text{MoS}_2$ top valence band and the Haldane model.
The difference is mostly concentrated in the $K/K'$ valley for both cases.
One potential reason for this phenomenon is that the Berry curvature itself is peaked in the two valleys.

\subsection{Construction of the Haldane Model}\label{sec:app-topo-haldane}

We construct the Haldane model \cite{haldane1988model} from the maximally localized Wannier functions \cite{Marzari1997maximally, Souza2001maximally} of graphene.
First, the ground state density is obtained through density functional theory calculations with a $18\times 18\times 1$ k-grid with a 50 Ry wavefunction cutoff via Quantum ESPRESSO \cite{Giannozzi2009quantum, Giannozzi2017advanced}.
After that, the Kohn-Sham wavefunctions and energies are obtained on a $50\times 50\times 1$ k-grid in the reciprocal space.
Two $p_z$-orbital-like Wannier functions are constructed on this grid using Wannier90 \cite{Pizzi2020wannier90}.
Subsequently, the nearest- and next-nearest-neighbor hoppings between the Wannier functions are extracted, and a phase $e^{\pm i\pi/2}$ is multiplied to the next-nearest-neighbor hopping to break the time-reversal symmetry, producing a two-band Haldane model \cite{haldane1988model}.
The two-band Hamiltonian is then diagonalized on the original $50\times 50$ k-grid.
The resulting eigenvectors, together with the gauge transformation matrices obtained from Wannierization, are then used to form linear combinations of the original Kohn-Sham Bloch states to obtain the real-space eigenfunctions of the Haldane model.

\section{Additional Results of Multilayer Perceptron} \label{sec:app-mlp}

\begin{figure}[!htbp]
    \centering
    \includegraphics[width=1.0\linewidth]{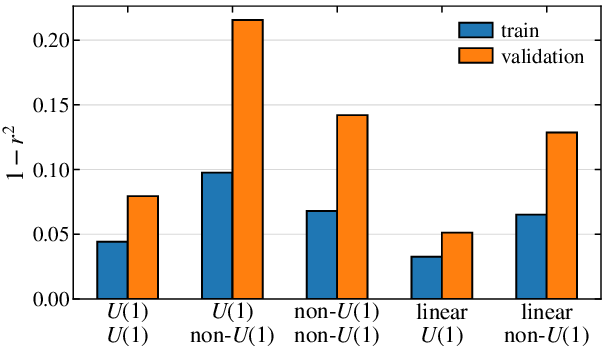}
    \caption{Difference between 1 and the $r^2$ score of different AE and MLP predictions. The first horizontal row of labels gives the AE version: $U(1)$-equivariant AE ($U(1)$), non-$U(1)$-equivariant AE (non-$U(1)$) and linear AE (linear). The second row gives the MLP version: $U(1)$-equivariant MLP ($U(1)$) and non-$U(1)$-equivariant MLP (non-$U(1)$).}
    \label{fig:r2bar}
\end{figure}

Fig. \ref{fig:r2bar} shows the gap between 1 and the $r^2$ score of each prediction model on the training and validation sets.
The height of each bar $1-r^2$ is positively related to the prediction error.
Though all models exhibit overfitting, the distinction between the models is notable, and the models considering $U(1)$ equivariance perform better than their non-equivariant counterparts.


%

\bibliography{ref}

\end{document}